\documentclass[12pt]{article}
\usepackage{exscale,epsfig,a4}
\begin{document}
\begin {titlepage}
\begin{flushleft}
FSUJ TPI QO-8/99
\end{flushleft}
\begin{flushright}
October, 1999
\end{flushright}
\vspace{2cm}

\begin{center}
{\Large \bf Measuring quantum state overlaps \\of traveling optical fields} 

\vspace{1.5cm}

 {\large \bf J. Clausen$^\mathrm{a}$\footnote{clausen@tpi.uni-jena.de}, 
  M. Dakna$^\mathrm{b}$, 
  L. Kn\"oll$^\mathrm{a}$ and D.--G. Welsch$^\mathrm{a}$}\\[1.ex]

\vspace{0.5cm}

{\large $^\mathrm{a}$ Friedrich-Schiller-Universit\"at Jena}\\[0.5 ex]
{\large Theoretisch-Physikalisches Institut}\\[0.5 ex]
{\large Max-Wien Platz 1, D-07743 Jena, Germany}

\vspace{0.5cm}

{\large $^\mathrm{b}$ Institut f\"{u}r Theoretische Physik,}\\[0.5 ex]
{\large Universit\"{a}t G\"{o}ttingen}\\[0.5 ex]
{\large Bunsenstr. 9, 73073 G\"{o}ttingen, Germany}

\vspace{2.5cm}
\end{center}
\begin{center}
\bf{Abstract}
\end{center}
We propose a detection scheme for measuring the overlap of the quantum state of 
a weakly excited traveling-field mode with a desired reference quantum state, 
by successive mixing the signal mode with modes prepared in coherent states and
performing photon-number measurements in an array of beam splitters. 
To illustrate the scheme, we discuss the measurement of the quantum phase
and the detection of Schr\"odinger-cat-like states. 
\end{titlepage}
\section{Introduction}
\label{sec1}
The problem of measuring the overlap of an unknown signal-field quantum state 
with a desired quantum state has been of increasing interest in quantum state 
measurement.
Typical setups that have been considered are photon chopping and related
photodetection schemes for measuring overlaps of the signal-field quantum state 
with Fock states \cite{Paul1,Mogilevtsev}, heterodyning for measuring overlaps 
of the signal-field quantum state with coherent states \cite{Shapiro1}, 
and balanced and unbalanced homodyning, respectively, for measuring overlaps of 
the signal-field quantum state with field-strength states \cite{Smithey1} and 
displaced Fock states \cite{Banaszek1}. 
It is therefore natural to ask the question of whether or not a universal setup 
can be constructed which can measure, in principle, arbitrary quantum 
mechanical overlaps. 

Surprisingly the answer is ´yes´. Extending earlier work on 
quantum state engineering \cite{Dakna1}, in this paper 
we show that arbitrary overlaps may be measured by feeding coherent states 
into a beam splitter array and performing zero and one photon measurements. 
Although in praxis the method cannot replace standard schemes such as 
homodyning for measuring field-strength distributions, it is noteworthy that 
it offers novel possibilities of measuring quantities whose direct detection 
has not been realized so far, particularly in a quantum regime. 
A well-known example is the problem of direct measurement of the quantum phase, 
where the overlaps of the signal quantum state with London phase states must be 
recorded.

The paper is organized as follows. In Section~\ref{sec2} the underlying
formalism is outlined and the basic formulas are given. Section~\ref{sec3} 
shows how an arbitrary overlap measurement may be performed and discusses
its efficiency. To illustrate the scheme, in Sections~\ref{sec4} and \ref{sec5} 
the measurement of the quantum phase statistics and  the statistics of 
Schr\"odinger-cat-like states are discussed respectively. 
A summary and some concluding remarks are given in Section~\ref{sec6}.
\section{Conditional quantum state transformation}
\label{sec2}
\begin{figure}[htp]
  \vspace{4cm}
  \unitlength1cm
  \begin{picture}(5,5)(7,27)
  \centering\psfig{figure=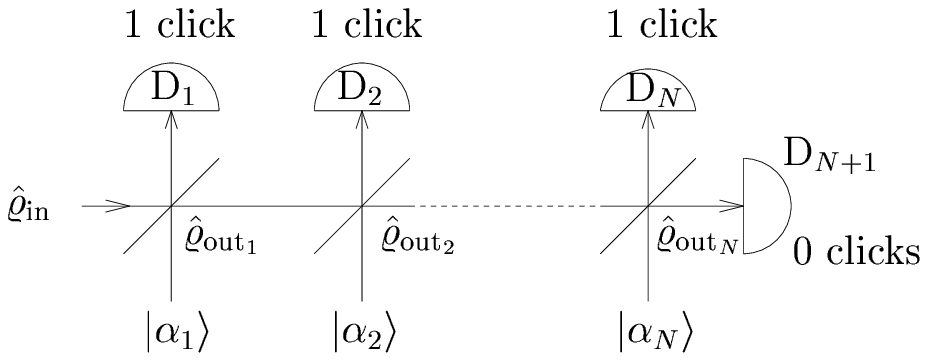,width=2\linewidth}
  \end{picture}
  \vspace{6cm}
\caption{
   Measurement of the overlap 
$\langle\Psi|\hat{\varrho}_{\mathrm{in}}|\Psi\rangle$ of a signal-quantum 
state $\hat{\varrho}_{\mathrm{in}}$ with a given state $|\Psi\rangle$ by 
successive mixing with modes prepared in appropriately chosen coherent states 
$|\alpha_1\rangle$ to $|\alpha_N\rangle$ at beam splitters and measuring 
the relative frequency of the event of detecting simultaneously 1 photon with 
the photodetectors $\mathrm{D}_1$ to $\mathrm{D}_N$ and 0 photons with 
$\mathrm{D}_{N+1}$.
\label{fig1}
}
\end{figure}

Let us consider the setup outlined in Fig.~\ref{fig1}. A signal-field mode 
prepared in a quantum state $\hat{\varrho}_{\mathrm{in}}$ passes an array of 
$N$ beam splitters at which it is mixed with modes prepared in coherent states 
$|\alpha_1\rangle$, $\!\ldots$, $\!|\alpha_N\rangle$, and photodetectors 
$\mathrm{D}_1$, $\!\ldots$, $\!\mathrm{D}_{N+1}$ measure photon numbers 
(clicks) of the outcoupled modes. 
   For the sake of simplicity, we assume that all beam splitters have the same 
transmittance $T$ and reflectance $R$. Further, we restrict our attention to 
perfect photodetection. 

We derive the transformation of the input quantum state step by step and begin 
with the first beam splitter and the associated photon-number measurement.
When the signal mode prepared in a quantum state $\hat{\varrho}_{\mathrm{in}}$
and a mode prepared in a coherent state $|\alpha_1\rangle$ are combined
and the detector $\mathrm{D}_1$ registers a photon, then 
$\hat{\varrho}_{\mathrm{in}}$ is transformed according to 
\begin{equation}
  \hat{\varrho}_{\mathrm{out}_1}=\frac{\hat{Y}_1\hat{\varrho}_{\mathrm{in}}
  \hat{Y}_1^\dagger}{p(1,1)} \,,
\label{-1}
\end{equation}
where
\begin{equation}
   p(1,1) 
   = {\mathrm{Tr}}\big(\hat{Y}_{1}
  \hat{\varrho}_{\mathrm{in}}\hat{Y}_{1}^\dagger\big) 
\label{0}
\end{equation}
is the probability of registering a photon, and
\begin{equation}
  \hat{Y}_{i} = -R^*\hat{D}\!\left(\frac{\alpha_{i}}{R^*}\right)
  T^{\hat{n}}\hat{a}
  \hat{D}\!\left(-\frac{T^*}{R^*}\alpha_{i}\right)
\label{1}
\end{equation}
[$\hat{D}(\alpha)$ $\!=$ 
$\!\mathrm{exp}(\alpha\hat{a}^\dagger$ $\!-$ $\!\alpha^*\hat{a})$]
is the nonunitary transformation operator for the conditional quantum-state 
transmission through the $i$th beam splitter \cite{Clausen1}. The output mode 
prepared in the quantum state $\hat{\varrho}_{\mathrm{out}_1}$ now enters the 
second beam splitter and is mixed with a mode prepared in a coherent state 
$|\alpha_2\rangle$. If the detector $\mathrm{D}_2$ associated with the second 
beam splitter also registers a photon, the transformed quantum state becomes 
\begin{equation}
  \hat{\varrho}_{\mathrm{out}_2}=\frac{\hat{Y}_2\hat{\varrho}_{\mathrm{out}_1}
  \hat{Y}_2^\dagger}{p(2,1|1,1)}\,,
\label{2}
\end{equation}
were 
\begin{equation}
   p(2,1|1,1)={\mathrm{Tr}}\big(\hat{Y}_2\hat{\varrho}_{\mathrm{out}_1}
   \hat{Y}_2^\dagger\big)
\label{3}
\end{equation}
is the probability of registering the second photon conditioned by the 
detection of the first photon. 
Using Eq.~(\ref{-1}), Eq.~(\ref{2}) can be rewritten as 
\begin{equation}
  \hat{\varrho}_{\mathrm{out}_2}=\frac{\hat{Y}_2\hat{Y}_1
  \hat{\varrho}_{\mathrm{in}}
  \hat{Y}_1^\dagger\hat{Y}_2^\dagger}{p(1,1;2,1)}
\label{4}
\end{equation}
were 
\begin{equation}
   p(1,1;2,1)={\mathrm{Tr}}\big(\hat{Y}_2\hat{Y}_1\hat{\varrho}_{\mathrm{in}}
   \hat{Y}_1^\dagger\hat{Y}_2^\dagger\big)
   = p(2,1|1,1)p(1,1)
\label{5}
\end{equation}
is the joint probability that the detectors $\mathrm{D}_1$ and $\mathrm{D}_2$ 
register a photon each. Repeating the procedure $N$ times, we find that the 
overall output quantum state is  
\begin{equation}
  \hat{\varrho}_{\mathrm{out}_N}=\frac{\hat{Y}\hat{\varrho}_{\mathrm{in}}
  \hat{Y}^\dagger}{p(1,1;2,1;\ldots;N,1)}\,,
\label{6}
\end{equation}
were 
\begin{equation}
  p(1,1;2,1;\ldots;N,1) = {\mathrm{Tr}}\big(\hat{Y}
  \hat{\varrho}_{\mathrm{in}}\hat{Y}^\dagger\big) 
\label{7}
\end{equation}
is the joint probability that all the detectors register a photon each. 
Obviously, the overall nonunitary transformation operator $\hat{Y}$ 
is given by the product of the operators $\hat{Y}_{i}$,
\begin{equation}
  \hat{Y}=\hat{Y}_N\cdots\hat{Y}_2\hat{Y}_1.
\label{8}
\end{equation}
\section{Measuring arbitrary overlaps}
\label{sec3}
Next let us consider a photon number measurement performed on the 
transformed signal quantum state $\hat{\varrho}_{\mathrm{out}_N}$
(detector $\mathrm{D}_{N+1}$ in Fig.~\ref{fig1}). The probability of 
detecting no photons is $\langle 0|\hat{\varrho}_{\mathrm{out}_N}|0\rangle$.
Equivalently, it is the probability of detecting no photons in the 
$(N+1)$th measurement conditioned by the detection of a photon in 
each of the $N$ preceding measurements,
\begin{equation}
  p(N+1,0|1,1;2,1;\ldots;N,1) 
  = \langle 0|\hat{\varrho}_{\mathrm{out}_N}|0\rangle.
\label{9}
\end{equation}
Combining Eqs.~(\ref{6}) and (\ref{9}), we easily see that 
\begin{equation}
  p(1,1;2,1;\ldots;N,1;N+1,0)=
  \langle0|\hat{Y}\hat{\varrho}_{\mathrm{in}}\hat{Y}^\dagger|0\rangle
\label{10}  
\end{equation}
is the joint probability that each of the detectors 
$\mathrm{D}_{1}$, $\!\ldots$, $\!\mathrm{D}_{N}$ registers a photon and the 
detector $\mathrm{D}_{N+1}$ does not register a photon. 

Let 
\begin{equation}
  \langle\Psi|\hat{\varrho}_{\mathrm{in}}|\Psi\rangle
  = {\mathrm{Tr}}\big(\hat{\varrho}_{\mathrm{in}}
  |\Psi\rangle\langle\Psi|\big)
\label{11}  
\end{equation}
be an overlap that is desired to be measured. When the values 
of the $N$ coherent amplitues $\alpha_{i}$ are chosen such that
\begin{equation}
  \frac{\hat{Y}^\dagger|0\rangle\langle0|\hat{Y}}
  {\|\hat{Y}^\dagger|0\rangle\|^2}=|\Psi\rangle\langle\Psi| 
\label{12}
\end{equation}
$\big(\||\Phi\rangle\|$ $\!=$ $\!\sqrt{\langle\Phi|\Phi\rangle}\big)$,
then the joint probability $p(1,1;2,1;\ldots;N,1;N+1,0)$ becomes 
proportional to $\langle\Psi|\hat{\varrho}_{\mathrm{in}}|\Psi\rangle$,
as is easily seen from Eq.~(\ref{10}),
\begin{equation}
  \langle\Psi|\hat{\varrho}_{\mathrm{in}}|\Psi\rangle=
  \frac{p(1,1;2,1;\ldots;N,1;N+1,0)}{\|\hat{Y}^\dagger
  |0\rangle\|^2}\,.
\label{13}
\end{equation}
In particular, performing the measurements for a set of states $|\Psi_l\rangle$ 
which can be used to define a positive operator valued measure (POVM) 
$\hat{\Pi}_l$ $\!=$ $\!|\Psi_l\rangle\langle\Psi_l|$, 
$\!l$ $\!=$ $\!1,2,\ldots$, directly yields the statistics of the quantity 
that is behind the POVM.

We now show that Eq.~(\ref{12}) can always be satisfied if the state 
$|\Psi\rangle$ is a finite superposition of Fock states, 
\begin{equation}
  |\Psi\rangle=\sum_{n=0}^N|n\rangle\langle n|\Psi\rangle.
\label{14}
\end{equation}
Note that the expansion of any physical state in the Fock basis can always be 
approximated to any desired degree of accuracy by truncating it at $N$ if $N$ 
is suitably large. The state $|\Psi\rangle$ in Eq.~(\ref{14}) is completely 
determined, e.g., by the $N$ zeros of its $Q$-function, i.e., the $N$ solutions 
$\beta_1$, $\!\ldots$, $\!\beta_N$ of the equation 
\begin{equation}
  \langle\Psi|\beta\rangle = 0 ,
\label{15}
\end{equation}
because of 
\begin{equation}
  |\Psi\rangle=\frac{\langle N|\Psi\rangle}{\sqrt{N!}}\prod_{k=1}^N
  (\hat{a}^\dagger-\beta_k^*)|0\rangle.
\label{16}
\end{equation}
In order to compare $|\Psi\rangle$ with $\hat{Y}^\dagger|0\rangle$, 
we use Eqs.~(\ref{1}) and (\ref{8}) and obtain
\begin{eqnarray}
  \lefteqn{
  \hat{Y}^\dagger|0\rangle = \mathrm{e}^{\mathrm{i}\xi}
  R^N \hat{D}\!\left(\frac{T^*}{R^*}\alpha_1
  \right)\hat{a}^\dagger {T^*}^{\hat{n}}
  \hat{D}\!\left(\frac{T^*\alpha_2\!-\!\alpha_1}{R^*}\right)
  \hat{a}^\dagger {T^*}^{\hat{n}}
  \hat{D}\!\left(\frac{T^*\alpha_3\!-\!\alpha_2}{R^*}\right)
  }  
  \nonumber\\ && \hspace{10ex} \times\;
  \hat{a}^\dagger {T^*}^{\hat{n}}\cdots
  \hat{D}\!\left(\frac{T^*\alpha_N\!-\!\alpha_{N-1}}{R^*}\right)
  \hat{a}^\dagger {T^*}^{\hat{n}}
  \hat{D}\!\left(-\frac{\alpha_N}{R^*}\right)
  |0\rangle,
\label{17}
\end{eqnarray}
where $\mathrm{e}^{\mathrm{i}\xi}$ is an irrelevant phase factor.
Now we rearrange the operator order such that the photon creation operators 
are on the left of the exponential operators, applying the rules 
$\hat{D}(\beta)\hat{a}^\dagger$ $\!=$ 
$\!(\hat{a}^\dagger$ $\!-$ $\!\beta^*)\hat{D}(\beta)$ and 
${T^*}^{\hat{n}}\hat{a}^\dagger$ $\!=$ $\!T^*\hat{a}^\dagger{T^*}^{\hat{n}}$.
After some calculation we derive  
\begin{equation}
  \hat{Y}^\dagger|0\rangle =  
  \frac{\mathrm{e}^{\mathrm{i}\xi^{\prime}} R^N}{T^{N(1-N)/2}}\,
  {\mathrm{exp}}\!\left(-\frac{1}{2}\sum_{k=1}^N|\alpha_k|^2\right)
  \!\prod_{k=1}^N \!\left(\hat{a}^\dagger\!-\!\frac{T^*}{R}\sum_{l=1}^k
  \!\frac{T\alpha_l^*\!-\!\alpha_{l-1}^*}{{T^*}^l}\right)\!|0\rangle
\label{18}
\end{equation}
($\alpha_0$ $\!=$ $\!0$). From a comparison of Eq.~(\ref{18}) with 
Eq.~(\ref{16}) it is seen that when the parameters $\alpha_k$ and $\beta_k$, 
$k$ $\!=$ $\!1,\ldots,N$, are related to each other as      
\begin{equation}
  \beta_k=\frac{T}{R^*}\sum_{l=1}^k\frac{T^*\alpha_l-\alpha_{l-1}}{T^l}\,,
\label{19}
\end{equation}
or equivalently, 
\begin{equation}
  \alpha_k=\frac{R^*}{T (T^*)^{k+1}}\sum_{l=1}^k|T|^{2l}(\beta_l-\beta_{l-1})
\label{20}
\end{equation}
($\beta_0$ $\!=$ $\!0$), then Eq.~(\ref{12}) is satisfied, i.e., the desired 
overlap is observed.

   From Eq.~(\ref{13}) it is seen that the sought overlap 
$\langle\Psi|\hat{\varrho}_{\rm in}|\Psi\rangle$ is determined by the measured
joint probability $p(1,1;2,1;\ldots;N,1;N+1,0)$ up to a factor of 
$\|\hat{Y}^\dagger|0\rangle\|^2$ which may be viewed, in a sense, as a measure 
of the efficiency of the detection scheme. Insertion of Eqs.~(\ref{18}) and
(\ref{16}) in Eq.~(\ref{12}) gives 
\begin{equation}
  \|\hat{Y}^\dagger|0\rangle\|^2 = \frac{N!}{|\langle N|\Psi\rangle|^2}\,
  |R|^{2N}|T|^{N(N-1)} \,
  \mathrm{exp}\!\left(-\sum_{k=1}^N|\alpha_k|^2\right).
\label{21}
\end{equation}
In particular, when the signal coincides with the state $|\Psi\rangle$,
i.e., $\hat{\varrho}_{\mathrm{in}}$ $\!=$ $\!|\Psi\rangle\langle\Psi|$,
then Eq.~(\ref{13}) reduces to 
$\|\hat{Y}^\dagger|0\rangle\|^2$ $\!=$ $\!p(1,1;2,1;\ldots;N,1;N+1,0)$.
Obviously, $\|\hat{Y}^\dagger|0\rangle\|^2$ is the joint probability of 
detecting one photon in the output channels $1,\ldots,N$ and no photon in the 
$(N$ $\!+$ $\!1)$th output channel in the case when the signal is prepared just 
in the state with which the overlap is desired to be measured. It therefore 
follows that $0$ $\!\le$ $\!\|\hat{Y}^\dagger|0\rangle\|^2\le1$. 
With increasing value of $N$ the value of $\|\hat{Y}^\dagger|0\rangle\|^2$ 
decreases rapidly in general and, accordingly, the number of recorded events 
must be increased in order to preserve accuracy. Note that 
$p(1,1;2,1;\ldots;N,1;N+1,0)$ $\!\leq$ $\!\|\hat{Y}^\dagger|0\rangle\|^2$, 
because of $\langle\Psi|\hat{\varrho}_{\rm in}|\Psi\rangle$ $\!\leq$ $\!1$ in 
Eq.~(\ref{13}).  
So far we have considered equal beam splitters which, for chosen state 
$|\Psi\rangle$, do not realize the maximally attainable efficiency in general. 
Clearly, the beam splitter parameters can individually specified such that 
the efficiency is maximized, depending on the state $|\Psi\rangle$.
\section{Quantum phase statistics}
\label{sec4}
In order to illustrate the scheme, let us first consider the problem of 
measuring the canonical phase statistics, i.e., the problem of measuring 
overlaps of a signal-mode quantum state with London phase states. 
So far, the most direct method for measuring the canonical phase has 
been direct sampling in balanced homodyning of the exponential moments 
of the canonical phase, i.e, the Fourier components of the canonical phase 
distribution \cite{Dakna-Breitenbach-1998}. To measure the canonical phase 
statistics directly, it was proposed to combine the signal mode with a 
reference mode prepared such that the measured output becomes proportional to 
the overlap between the signal quantum state and phase states \cite{Barnett1}.
However, the method also called projection synthesis requires 
the reference mode to be prepared in highly involved nonclassical states.

\begin{figure}[htp]
  \vspace{4cm}
  \unitlength1cm
  \begin{picture}(5,5)(5.7,27)
  \centering\psfig{figure=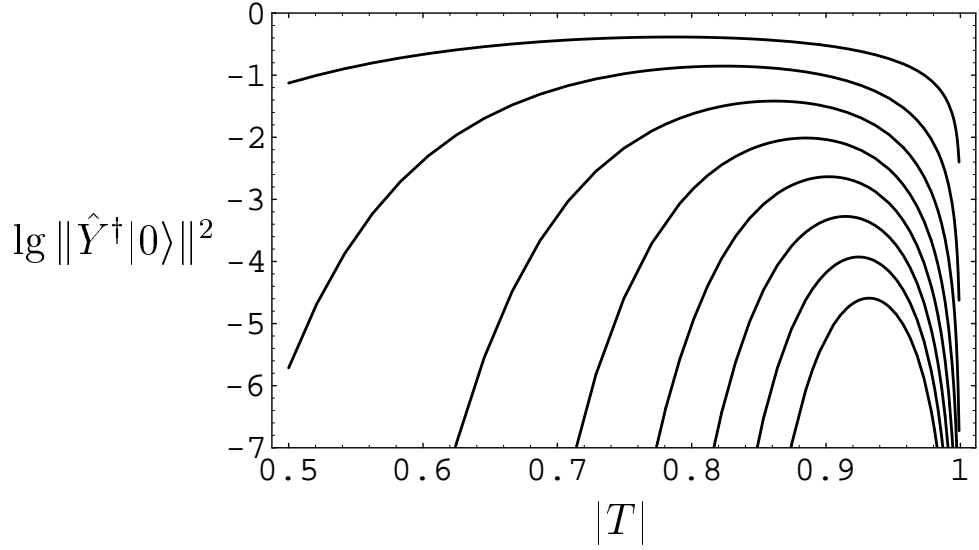,width=2\linewidth}
  \end{picture}
  \vspace{6cm}
\caption{
Efficiency $\|\hat{Y}^\dagger|0\rangle\|^2$, Eq.~(\protect\ref{21}), 
for measuring the overlap of a signal-mode quantum state with truncated 
London phase states $|\varphi;N\rangle$, Eq.~(\protect\ref{22}), as a 
function of the absolute value of the beam-splitter transmittance $|T|$ 
for $\varphi$ $\!=$ $\!0$, the value of $N$ being varied from
$N$ $\!=$ $1\!$ (top curve) to $N$ $\!=$ $\!8$ (bottom curve). 
\label{fig2}
}
\end{figure}

On the contrary, our method only uses coherent states, without need to 
explicitly excite nonclassical states, so that it can be applied employing 
nowadays available techniques. Since any (physical) quantum state can be 
truncated at some photon number $N$ in the Fock basis, it is sufficient to 
consider the overlap with truncated London phase states,
\begin{equation}
  |\varphi;N\rangle=\frac{1}{\sqrt{N+1}}\sum_{n=0}^{N}
  \mathrm{e}^{\mathrm{i}n\varphi}\,|n\rangle.
\label{22}
\end{equation}
The efficiency $\|\hat{Y}^\dagger|0\rangle\|^2$, Eq.~(\ref{21}), for 
measuring overlaps with these states is plotted in Fig.~\ref{fig2} for 
$N$ $\!=$ $\!1,\ldots,8$. It is seen that the efficiency rapidly decreases 
with increasing $N$. Hence, the signal field should contain only few photons 
in practice. This is just the most interesting case for studying 
the quantum phase, otherwise the phase behaves nearly classically. 

In the simple case when the signal-mode quantum state reduces to a (coherent) 
superposition of the vacuum state $|0\rangle$ and a one-photon Fock state 
$|1\rangle$, we have
\begin{equation}
  |z\rangle =  \frac{1}{\sqrt{1+|z|^2}}(|0\rangle + z |1\rangle).
\end{equation}
The overlap between the signal-mode quantum state 
$\hat{\varrho}_{\rm in}$ $\!=$ $\!|z\rangle\langle z|$ and a state 
$|\Psi\rangle$ $\!=$ $|\varphi;N\rangle$ then reads
\begin{equation}
  \langle\Psi|\hat{\varrho}_{\mathrm{in}}|\Psi\rangle = 
  |\langle z|\varphi;N\rangle|^2 = 
  \frac{1+|z|^2+2|z|\cos(\varphi-\psi)}{(N+1)(1+|z|^2)}
\label{23}  
\end{equation}
where $z$ $\!=$ $\!|z|e^{i\psi}$. Measuring the overlap for all phases 
$\varphi$ yields the canonical phase distribution 
\begin{equation}
  \mbox{prob}\,\varphi = \frac{N+1}{2\pi}\, 
  \langle\Psi|\hat{\varrho}_{\mathrm{in}}|\Psi\rangle.  
\label{24}  
\end{equation}
Note that the right-hand side in Eq.~(\ref{24}) is independent of $N$,
so that only one beam splitter in Fig.~\ref{fig1} is required. 
The canonical phase distribution is then given, according to Eq.~(\ref{13}), 
by the measured two-event joint probabilities $p(1,1;2,0)$,   
\begin{equation}
  \mbox{prob}\,\varphi = \frac{1}{\pi}\, 
  \frac{p(1,1;2,0)}{\|\hat{Y}^\dagger|0\rangle\|^2}.
\label{25}  
\end{equation}
   From Eq.~(\ref{20}) it is easily seen that the values of the coherent-state 
amplitude must be chosen such that 
$\alpha$ $\!=$ $\!-(R^*/T^*)\mathrm{e}^{\mathrm{i}\varphi}$,   
and from Eq.~(\ref{21}) it follows that
\begin{equation}
  \|\hat{Y}^\dagger|0\rangle\|^2 = 2|R|^2 \mathrm{e}^{-|R/T|^2}.
\label{26}
\end{equation}
We see that $\|\hat{Y}^\dagger|0\rangle\|^2$ attains at $|T|^2$ $\!=$ $\!0.62$
a maximum of $\|\hat{Y}^\dagger|0\rangle\|^2_{{\rm max}}$ $\!=$ $\!0.41$.
Obviously, in the limit when $|z|$ $\!\to$ $\!0$ then the canonical phase of
the vacuum is detected.

\begin{figure}[htp]
  \vspace{4cm}
  \unitlength1cm
  \begin{picture}(5,5)(6.2,27)
  \centering\psfig{figure=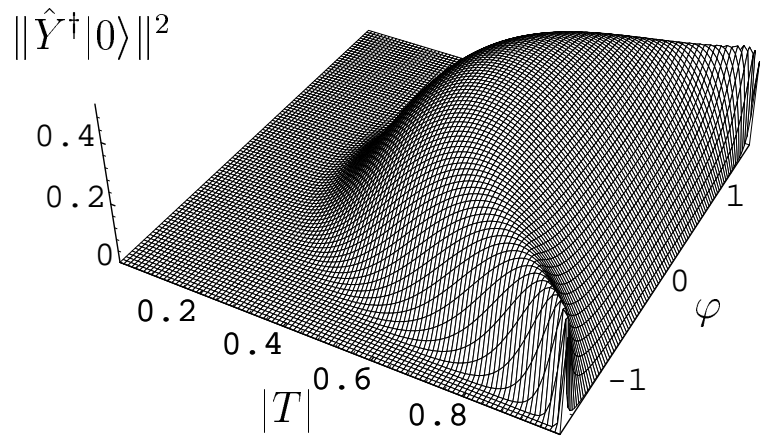,width=2\linewidth}
  \end{picture}
  \vspace{6cm}
\caption{
Efficiency $\|\hat{Y}^\dagger|0\rangle\|^2$, Eq.~(\protect\ref{30}), 
for measuring the overlap of a signal-mode quantum state with a state 
$|\varphi,\chi;1\rangle$, Eq.~(\protect\ref{27}), as a function of the phase 
$\varphi$ and the absolute value of the beam-splitter transmittance $|T|$.
\label{fig3}
}
\end{figure}

\begin{figure}[htp]
  \vspace{4cm}
  \unitlength1cm
  \begin{picture}(5,5)(6.5,27)
  \centering\psfig{figure=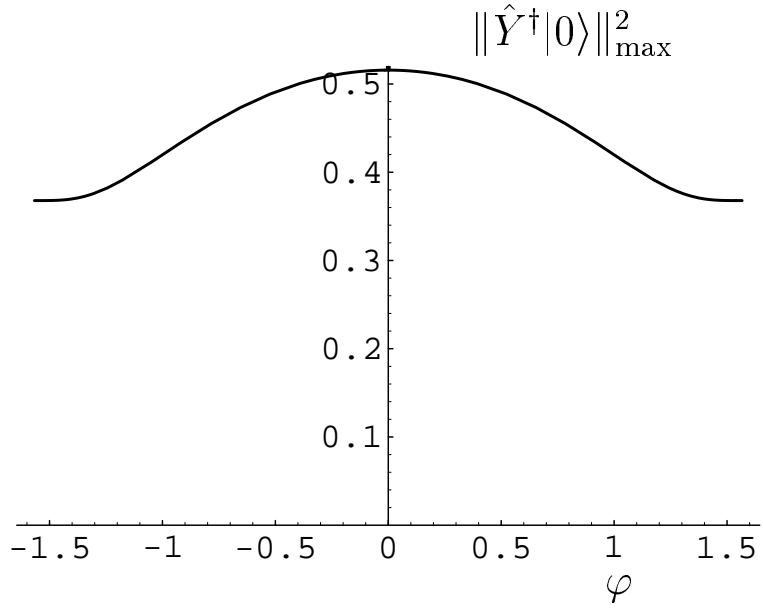,width=2\linewidth}
  \end{picture}
  \vspace{6cm}
\caption{
   Maximal efficiency $\|\hat{Y}^\dagger|0\rangle\|^2_{\mathrm{max}}$ 
as a function of the phase $\varphi$ as obtained with Eq.~(\protect\ref{30}) 
if $|T|$ is chosen according to Eq.~(\protect\ref{32}).
\label{fig4}
}
\end{figure}

The method can also be used to measure the phase distributions that correspond 
to the Hermitian cosine- and sine-phase operators. 
In particular, the method allows one, for the first time, to measure the 
nontrivial cosine- and sine-phase distributions of the vacuum directly. 
The (truncated) cosine- and sine-phase states are superpositions of two 
(truncated) London phase states each (for details, see \cite{VogelWelsch}),  
\begin{equation}
  |\varphi,\chi;N\rangle=C(\varphi;N) 
  \left[\mathrm{e}^{\mathrm{i}\varphi} |\chi+\varphi;N\rangle - 
  \mathrm{e}^{-\mathrm{i}\varphi} |\chi-\varphi;N\rangle\right],
\label{27}
\end{equation}
with $C(\varphi;N)$ 
$\!=$ $\!-\mathrm{i}\{2$ $\!-$ $\!2\sin(N$ $\!+$ $\!1)\varphi\cos(N$ 
$\!+$ $\!2)\varphi/[(N$ $\!+$ $\!1)\sin\varphi]\}^{-1/2}$.
In particular, 
$|\varphi,\chi$ $\!=$ $\!0;N\rangle$ $\!=$ $\!|\cos\varphi;N\rangle$ and 
$|\varphi,\chi$ $\!=$ $\!\frac{\pi}{2};N\rangle$ $\!=$ 
$\!|\sin(\frac{\pi}{2}$ $\!-$ $\!\varphi);N\rangle$
are the (truncated) Susskind-Glogower cosine- and sine-phase states, 
respectively. The overlap between the signal-mode quantum state 
$\hat{\varrho}_{\rm in}$ $\!=$ $\!|z\rangle\langle z|$ and a state 
$|\Psi\rangle$ $\!=$ $|\varphi,\chi;N\rangle$ reads
\begin{eqnarray}
  \lefteqn{
  \langle\Psi|\hat{\varrho}_{\mathrm{in}}|\Psi\rangle 
  = |\langle z|\varphi,\chi;N\rangle|^2
  = \frac{4|C(\varphi;N)|^2}{(N+1)(1+|z|^2)}
  }
  \nonumber\\&&\hspace{12ex}\times\;  
  \left[\sin^2\varphi+2|z|\cos(\psi\!-\!\chi)
  \sin\varphi\sin(2\varphi)+|z|^2\sin^22\varphi\right]. 
  \qquad  
\label{28}
\end{eqnarray}
Measuring this overlap for all phases $\varphi$ yields, for chosen $\chi$, 
the corresponding Susskind-Glogower trigonometric phase distribution  
\begin{equation}
  \left.\mbox{prob}\,\varphi\right|_{\chi} 
  = \frac{N+1}{2\pi|C(\varphi;N)|^2}\, 
  \langle\Psi|\hat{\varrho}_{\mathrm{in}}|\Psi\rangle.
\label{29}
\end{equation}
Again, the right-hand side in Eq.~(\ref{29}) is independent of $N$, so that 
it is sufficient to measure the joint probabilities for $N$ $\!=$ $\!1$, i.e., 
$p(1,1;2,0)$. From Eq.~(\ref{20}) the values of the coherent-state amplitude 
are then found to be
$\alpha=-[R^*/(2T^*\cos\varphi)]\mathrm{e}^{\mathrm{i}\chi}$,
and Eq.~(\ref{21}) yields the efficiency
\begin{equation}
  \|\hat{Y}^\dagger|0\rangle\|^2 =
  \frac{1-\cos\varphi\cos(3\varphi)}{\sin^2(2\varphi)}\,|R|^2
  \mathrm{e}^{-|R/(2T\cos\varphi)|^2}
\label{30}
\end{equation}
which is plotted in Fig.~\ref{fig3}.
Combining Eqs.~(\ref{13}), (\ref{29}), and (\ref{30}), we obtain 
\begin{equation}
  \left.\mbox{prob}\,\varphi\right|_{\chi} 
  = \frac{2 \sin^2(2\varphi)}{\pi|R|^2}\,\mathrm{e}^{|R/(2T\cos\varphi)|^2}
  p(1,1;2,0).
\label{31}
\end{equation}
   From Eq.~(\ref{30}) it can be found that $\|\hat{Y}^\dagger|0\rangle\|^2$ 
becomes maximal for 
\begin{equation}
  |T|^2 = \frac{\sqrt{1+16\cos^2\varphi}-1}{8\cos^2\varphi}
\label{32}
\end{equation}
[cf. Fig.~\ref{fig4}], and hence
$0.36$ $\!<$ $\|\hat{Y}^\dagger|0\rangle\|^2_{\rm max}$ $\!<$ $\!0.52$ for 
$\varphi$ within a $\pi$ interval. Note that $|\hat{Y}^\dagger|0\rangle\|^2$ 
does not depend on $\chi$. In the limit when $|z|$ $\!\to$ $\!0$, then the 
nonuniformly distributed Susskind-Glogower trigonometric phase of the vacuum 
is detected.
\section{Schr\"odinger-cat state statistics}
\label{sec5}
Let us return to the general scheme in Fig.~\ref{fig1}. 
When some of the zeros $\beta_k^*$ in Eq.~(\ref{16}) are equal, 
\begin{equation}
  |\Psi\rangle=\frac{\langle N|\Psi\rangle}{\sqrt{N!}}\prod_{l=1}^M
  (\hat{a}^\dagger-\beta_l^*)^{d_l}|0\rangle,
\label{40}  
\end{equation}
($M$ $\!<N$ $\!$), then the number of detectors can be reduced to 
$M$ $\!+$ $\!1$, and the overlap is given by 
$p(1,d_1;$ $\!\ldots;$ $\!M,d_M;M+1,0)$. It can be easily proved that the 
nonunitary transformation operator $\hat{Y}$ now reads 
\begin{equation}
  \hat{Y} = \hat{Y}_M\ldots\hat{Y}_2\hat{Y}_1, 
\label{41}  
\end{equation}
where $\hat{Y}_{i}$ is given by Eq.~(\ref{1}), with 
$(-R^*\hat{a})^{d_i}/\sqrt{d_i!}$ in place of $-R^*\hat{a}$, 
and the overlap $\langle\Psi|\hat{\varrho}_{\rm in}|\Psi\rangle$ is given by 
\begin{equation}
  \langle\Psi|\hat{\varrho}_{\mathrm{in}}|\Psi\rangle=
  \frac{p(1,d_1;2,d_2;\ldots;M,d_M;M+1,0)}{\|\hat{Y}^\dagger
  |0\rangle\|^2}
\label{42}
\end{equation}
in place of Eq.~(\ref{13}).

An interesting example is the measurement of the overlap 
of a signal-mode quantum state with quantum states  
\begin{equation}
  |\Psi_n(\beta)\rangle={\cal N}^{-1/2}
  \big[(\hat{a}^\dagger)^2-(\beta^*)^2\big]^n\,|0\rangle.
\label{43}
\end{equation}
Applying standard formulas \cite{Prudnikov}, the normalization factor 
${\cal N}$ can be expressed in terms of the generalized hypergeometric 
function $_1{\rm F}_2(a,b,c;z)$,
\begin{equation}
  {\cal N}=\sum_{k=0}^n{n \choose k}^2\frac{(2k)!}{|\beta|^{4(k-n)}} =
  \frac{4^n n!}{\sqrt{\pi}}\,
  \Gamma\!\left(n\!+\!{\textstyle\frac{1}{2}}\right)
  {_1{\rm F}_2}\!\left(-n,{\textstyle\frac{1}{2}}\!-\!n,1,
  {\textstyle\frac{1}{4}} |\beta|^4\right). 
\label{44}   
\end{equation}
   For $n$ $\!=$ $\!|\beta|^2$ and increasing $|\beta|^2$ the states
$|\Psi_n(\beta)\rangle$ behave like Schr\"odinger-cat states
\begin{equation}
  |\Psi(\beta)\rangle
  =\frac{\mathrm{e}^{\mathrm{i}|\beta|^2(\pi-2\varphi_\beta)}}{\sqrt{2}}
  \left(|\mathrm{i}\beta\rangle + |-\mathrm{i}\beta\rangle\right),
\label{45}
\end{equation}
as can be seen from the overlap $|\langle\Psi_n(\beta)|\Psi(\beta)\rangle|^2$ 
for $|\beta|^2$ $\!=$ $\!n$,
\begin{equation}
  |\langle\Psi_n(\beta)|\Psi(\beta)\rangle|^2 = 
  2\left(\frac{4}{\mathrm{e}}\right)^n\left(\sum_{k=0}^np_kf_k\right)^{-1},
\label{46}
\end{equation}
where
\begin{equation}
  p_k=\frac{1}{2^n}{n \choose k},
  \qquad
  f_k=2^n{n \choose k}\frac{(2k)!}{n^{2k}}.
\label{46a}
\end{equation}
   From the asymptotic behavior of $p_k$ it can be seen that for sufficiently 
large $n$ the sum in Eq.~(\ref{46}) can be replaced by $f_{\frac{n}{2}}$. 
Applying the Stirling formula to $f_{\frac{n}{2}}$ then yields 
$|\langle\Psi_n(\beta)|\Psi(\beta)\rangle|^2$ $\!\to$ $\!1$ if 
$n$ $\!\to$ $\!\infty$ in Eq.~(\ref{46}). Note that the phase factor 
$\mathrm{exp}[\mathrm{i}|\beta|^2(\pi-2\varphi_\beta)]$ in Eq.~(\ref{45}) 
follows directly from $\langle\mathrm{i}\beta|\Psi_n(\beta)\rangle$. 

\begin{figure}[htp]
  \vspace{4cm}
  \unitlength1cm
  \begin{picture}(5,5)(5.7,27)
  \centering\psfig{figure=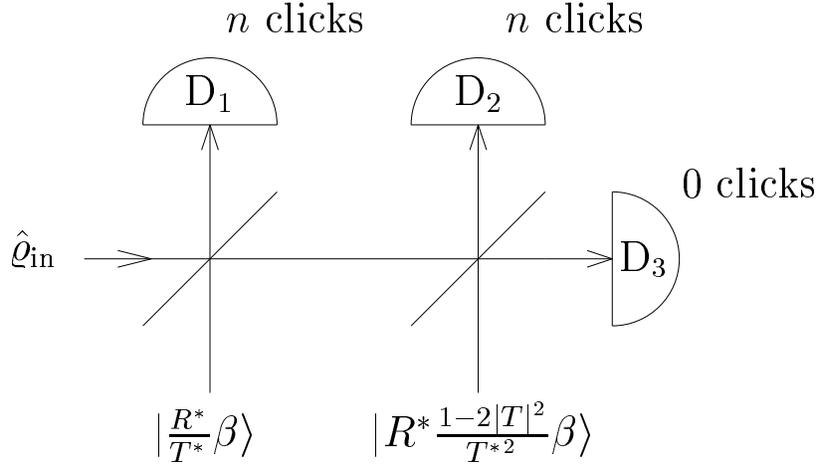,width=2\linewidth}
  \end{picture}
  \vspace{6cm}
\caption{
   Measurement of the overlap 
$\langle\Psi_n(\beta)|\hat{\varrho}_{\mathrm{in}}|\Psi_n(\beta)\rangle$ of a 
signal-quantum state $\hat{\varrho}_{\mathrm{in}}$ with a state 
$|\Psi_n(\beta)\rangle$, Eq.~(\protect\ref{43}), by mixing the signal mode with 
two modes prepared in coherent states at two beam splitters and measuring the 
relative frequency of the event of detecting simultaneously $n$ photons with 
the photodetectors $\mathrm{D}_1$ and $\mathrm{D}_2$ and $0$ photons with 
$\mathrm{D}_3$.
\label{fig5}
}
\end{figure}

\begin{figure}[htp]
  \vspace{4cm}
  \unitlength1cm
  \begin{picture}(5,5)(5.5,27)
  \centering\psfig{figure=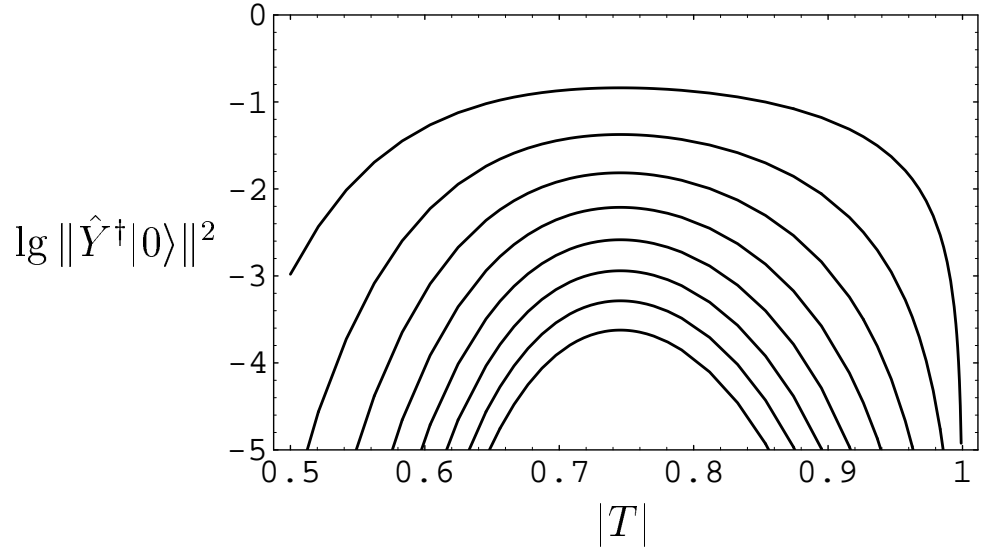,width=2\linewidth}
  \end{picture}
  \vspace{6cm}
\caption{
Efficiency $\|\hat{Y}^\dagger|0\rangle\|^2$, Eq.~(\protect\ref{21}), 
for measuring the overlap of a signal-mode quantum state with 
Schr\"{o}dinger-cat-like states $|\Psi_{n=|\beta|^2}(\beta)\rangle$, 
Eq.~(\protect\ref{43}), as a function of the absolute value of the 
beam-splitter transmittance $|T|$, the value of $n$ being varied from 
$n$ $\!=$ $1\!$ (top curve) to $n$ $\!=$ $\!8$ (bottom curve). 
\label{fig6}
}
\end{figure}

It is worth noting that $|\Psi_{n=|\beta|^2}(\beta)\rangle$ is already a good 
approximation of $|\Psi(\beta)\rangle$ for relatively small values of 
$|\beta|^2$, e.g.,  
$|\langle\Psi_{n=|\beta|^2}(\beta)|\Psi(\beta)\rangle|^2$ $\!$ $\!>0.95$ for 
\mbox{$|\beta|^2$ $\!\ge$ $\!3$}. 
Hence, measuring of the overlap of a signal-mode quantum state with a state 
$|\Psi_{n=|\beta|^2}(\beta)\rangle$ corresponds, in good approximation, to 
measuring the overlap with a Schr\"{o}dinger-cat state $|\Psi(\beta)\rangle$. 
The two-beam-splitter scheme that realizes the measurement is shown in 
   Fig.~\ref{fig5}. The efficiency of the measurement is given by
\begin{equation}
  \|\hat{Y}^\dagger|0\rangle\|^2=\frac{{\cal N}|R^2T|^{2n}}{n!^2}
  \,\exp\!\left\{-\left|\frac{R\beta}{T}\right|^2\left[1+
  |T|^{-2}\left(1-2|T|^2\right)^2\right]\right\}.
\label{47}
\end{equation}
Plots of $\|\hat{Y}^\dagger|0\rangle\|^2$ are shown in Fig.~\ref{fig6} for 
various values of $n$. 
The scheme can easily be extended in order to measure the overlap of 
$\hat{\varrho}_{\mathrm{in}}$ with an arbitrary superposition of 
two coherent states,
$|\Psi(\beta_1,\beta_2)\rangle$ $\!=$ $\!2^{-1/2}(|\beta_1\rangle$ 
$\!+$ $\!|\beta_2\rangle)$. For this purpose the quantum state of the signal 
mode must be coherently displaced, $\hat{\varrho}_{\rm in}$ $\!\to$ 
$\!\hat{D}^\dagger(\gamma)\hat{\varrho}_{\rm in}\hat{D}(\gamma)$,
before measuring the overlap with $|\Psi(\beta)\rangle$.
Choosing the parameters such that 
$\beta$ $\!=(\beta_1$ $\!-$ $\!\beta_2)/(2\mathrm{i})$
and $\gamma$ $\!=$ $\!(\beta_1$ $\!+$ $\!\beta_2)/2$, 
it is easily seen that
\begin{equation}
  \langle\Psi(\beta_1,\beta_2)|\hat{\varrho}_{\rm in}
  |\Psi(\beta_1,\beta_2)\rangle
  = \langle\Psi(\beta)
  |\hat{D}^\dagger(\gamma)\hat{\varrho}_{\rm in}\hat{D}(\gamma)
  |\Psi(\beta)\rangle.
\label{48}
\end{equation}
Note that a displacement may be achieved by mixing the signal with a mode 
prepared in a strong coherent state using a highly transmitting beam splitter 
(see, e.g, \cite{Paris1}).
\section{Conclusion}
\label{sec6}
We have presented a scheme for the direct measurement of the overlaps of an 
unknown signal quantum state with arbitrary reference quantum states. It is 
based on superimposing the signal mode and modes prepared in coherent states 
and detecting specific coincident events in the photon statistics of the 
outgoing modes. The advantage of the scheme is that the measurements can be 
performed without any explicit preparation of the states the signal-mode 
quantum state is projected onto. The disadavantage is that the relative 
frequency of the desired events can be very small in general, which is typical 
of conditional measurement schemes. The applicability of the method is 
therefore expected to be restricted to measuring overlaps in which only a few 
photons are involved, i.e., it is the highly nonclassical area around the 
quantum vacuum which is covered by the method.

In particular, the method may be used for measuring quantum overlaps for which 
a direct, dynamical measurement method has not been available so far. In this 
context, we have considered the measurement of the overlaps of a signal-mode 
quantum state with canonical phase states and related cosine- and  sine-phase 
states, particularly near the quantum vacuum, and calculated the detection 
probabilities. Further, we have considered the problem of measurement of the 
overlaps of a signal-mode quantum state with states that are superpositions of 
two coherent states. Recently it has been shown that from such overlaps the 
quantum state of the signal mode can be inferred \cite{Freyberger}. In 
principle, these overlaps could be measured, as proposed in \cite{Freyberger}, 
in heterodyne detection, however with the reference mode being prepared in a 
superposition of two coherent states in place of a single coherent state 
-- a rather difficult problem which has not been solved so far in practice.
\section*{Acknowledgement}
This work was supported by the Deutsche Forschungsgemeinschaft. 

\newpage
\newpage
\newpage
\newpage
\newpage
\newpage

\end{document}